\documentclass[conference]{IEEEtran}
\makeatletter\if@twocolumn\PassOptionsToPackage{switch}{lineno}\else\fi\makeatother

\usepackage{graphicx}
\usepackage{xurl}
\usepackage[T1]{fontenc}
\ifCLASSINFOpdf
\else
\fi
\usepackage{amsmath}
\usepackage[cmintegrals]{newtxmath}
\usepackage{url,multirow,morefloats,floatflt,cancel,tfrupee}
\makeatletter

\AtBeginDocument{\@ifpackageloaded{textcomp}{}{\usepackage{textcomp}}}
\makeatother
\usepackage{colortbl}
\usepackage{xcolor}
\usepackage{pifont}
\usepackage[nointegrals]{wasysym}
\makeatletter

\def\mcWidth#1{\csname TY@F#1\endcsname+\tabcolsep}

\def\cAlignHack{\rightskip\@flushglue\leftskip\@flushglue\parindent\z@\parfillskip\z@skip}
\def\rAlignHack{\rightskip\z@skip\leftskip\@flushglue \parindent\z@\parfillskip\z@skip}

\@ifundefined{etal}{}{}

\usepackage{ifxetex}
\ifxetex\else\if@twocolumn\@ifpackageloaded{stfloats}{}{\usepackage{dblfloatfix}}\fi\fi

\AtBeginDocument{
\expandafter\ifx\csname eqalign\endcsname\relax
\def\eqalign#1{\null\vcenter{\def\\{\cr}\openup\jot\m@th
  \ialign{\strut$\displaystyle{##}$\hfil&$\displaystyle{{}##}$\hfil
      \crcr#1\crcr}}\,}
\fi
}

\AtBeginDocument{%
  \@ifpackageloaded{endfloat}%
   {\renewcommand\efloat@iwrite[1]{\immediate\expandafter\protected@write\csname efloat@post#1\endcsname{}}}{\newif\ifefloat@tables}%
}%

\def\BreakURLText#1{\@tfor\brk@tempa:=#1\do{\brk@tempa\hskip0pt}}
\let\lt=<
\let\gt=>
\def\processVert{\ifmmode|\else\textbar\fi}

\@ifundefined{subparagraph}{
\def\subparagraph{\@startsection{paragraph}{5}{2\parindent}{0ex plus 0.1ex minus 0.1ex}%
{0ex}{\normalfont\small\itshape}}%
}{}

\newcommand\role[1]{\unskip}
\newcommand\aucollab[1]{\unskip}
  
\@ifundefined{tsGraphicsScaleX}{\gdef\tsGraphicsScaleX{1}}{}
\@ifundefined{tsGraphicsScaleY}{\gdef\tsGraphicsScaleY{.9}}{}
\def\checkGraphicsWidth{\ifdim\Gin@nat@width>\linewidth
	\tsGraphicsScaleX\linewidth\else\Gin@nat@width\fi}

\def\checkGraphicsHeight{\ifdim\Gin@nat@height>.9\textheight
	\tsGraphicsScaleY\textheight\else\Gin@nat@height\fi}

\def\fixFloatSize#1{}
\let\ts@includegraphics\includegraphics

\def\inlinegraphic[#1]#2{{\edef\@tempa{#1}\edef\baseline@shift{\ifx\@tempa\@empty0\else#1\fi}\edef\tempZ{\the\numexpr(\numexpr(\baseline@shift*\f@size/100))}\protect\raisebox{\tempZ pt}{\ts@includegraphics{#2}}}}

\AtBeginDocument{\def\includegraphics{\@ifnextchar[{\ts@includegraphics}{\ts@includegraphics[width=\checkGraphicsWidth,height=\checkGraphicsHeight,keepaspectratio]}}}

\DeclareMathAlphabet{\mathpzc}{OT1}{pzc}{m}{it}

\def\URL#1#2{\@ifundefined{href}{#2}{\href{#1}{#2}}}

\def\UrlOrds{\do\*\do\-\do\~\do\'\do\"\do\-}%
\g@addto@macro{\UrlBreaks}{\UrlOrds}

\edef\fntEncoding{\f@encoding}

\makeatother

\newif\ifmultipleabstract\multipleabstractfalse%
\makeatletter
\AtBeginDocument{\@ifpackageloaded{longtable}{%
\def\LT@makecaption#1#2#3{%
  \LT@mcol\LT@cols c{\hbox to\z@{\hss\parbox[t]\LTcapwidth{%
    \sbox\@tempboxa{#1{#2: } #3}%
    \ifdim\wd\@tempboxa>\hsize
      #1{#2: }\textsc{#3}%
    \else
      \hbox to\hsize{\hfil\box\@tempboxa\hfil}%
    \fi
    \endgraf\vskip\baselineskip}%
  \hss}}}
}{}}
\makeatother

\makeatletter
\let\citep\cite
\let\citet\cite
\makeatother

   \makeatletter
  \def\fig@textbf{\textbf}
   \AtBeginDocument{\renewcommand\floatc@plain[2]{\setbox\@tempboxa\hbox{{\footnotesize#1.}\footnotesize\hskip.5em#2}%
    \ifdim\wd\@tempboxa>\hsize {\fig@textbf{\footnotesize#1.}}\footnotesize\hskip.5em#2\par
        \else\hbox to\hsize{\hfil\box\@tempboxa\hfil}\fi}}
    \makeatother

\usepackage{float}

\begin{document}

\nocite{*}

%


        \title{Performance Improvement of IaaS Type of Cloud Computing Using Virtualization Technique}
      \author{
		\IEEEauthorblockN{Dawit~Zeleke}\\[-12pt]Email: dawit.zeleke.aau@gmail.com ~\\(Corresponding author)
        \vspace*{1pc}\and 
		\IEEEauthorblockN{Admassu}}
  


\maketitle 

\begin{abstract}
\textbf{Abstract}\textbf{{\textemdash}Cloud}\textbf{\space }\textbf{computing}\textbf{\space }\textbf{has}\textbf{\space }\textbf{transformed}\textbf{\space }\textbf{the}\textbf{\space }\textbf{way}\textbf{\space }\textbf{orga-}\textbf{\space }\textbf{nizations manage and scale their IT infrastructure by offering}\textbf{\space }\textbf{flexible,}\textbf{\space }\textbf{scalable,}\textbf{\space }\textbf{and}\textbf{\space }\textbf{cost-effective}\textbf{\space }\textbf{solutions.}\textbf{\space }\textbf{However,}\textbf{\space }\textbf{the}\textbf{\space }\textbf{Infras-}\textbf{\space }\textbf{tructure as a Service (IaaS) model faces performance challenges}\textbf{\space }\textbf{primarily due to the limitations imposed by virtualization tech-}\textbf{\space }\textbf{nology.}\textbf{\space }\textbf{This}\textbf{\space }\textbf{paper}\textbf{\space }\textbf{focuses}\textbf{\space }\textbf{on}\textbf{\space }\textbf{designing}\textbf{\space }\textbf{an}\textbf{\space }\textbf{effective}\textbf{\space }\textbf{virtualization}\textbf{\space }\textbf{technique for IaaS, aiming to improve infrastructure-level per-}\textbf{\space }\textbf{formance. Through a systematic literature review and a design,}\textbf{\space }\textbf{development,}\textbf{\space }\textbf{and}\textbf{\space }\textbf{evaluation}\textbf{\space }\textbf{approach,}\textbf{\space }\textbf{various}\textbf{\space }\textbf{virtualization}\textbf{\space }\textbf{techniques}\textbf{\space }\textbf{such}\textbf{\space }\textbf{as}\textbf{\space }\textbf{full}\textbf{\space }\textbf{virtualization,}\textbf{\space }\textbf{paravirtualization,}\textbf{\space }\textbf{and}\textbf{\space }\textbf{hardware-assisted}\textbf{\space }\textbf{virtualization}\textbf{\space }\textbf{are}\textbf{\space }\textbf{explored.}\textbf{\space }\textbf{The}\textbf{\space }\textbf{study}\textbf{\space }\textbf{also}\textbf{\space }\textbf{considers the role of hypervisors like Xen, KVM, and VMware}\textbf{\space }\textbf{ESXi}\textbf{\space }\textbf{in}\textbf{\space }\textbf{improving}\textbf{\space }\textbf{performance.}\textbf{\space }\textbf{The}\textbf{\space }\textbf{proposed}\textbf{\space }\textbf{solution}\textbf{\space }\textbf{seeks}\textbf{\space }\textbf{to optimize resource utilization, minimize latency, and enhance}\textbf{\space }\textbf{overall throughput in IaaS environments. Finally, the research}\textbf{\space }\textbf{discusses the potential application of this virtualization technique}\textbf{\space }\textbf{for public cloud computing solutions tailored for Ethiopian Small}\textbf{\space }\textbf{and Medium Enterprises (ESMEs) using platforms like Amazon} \textbf{EC2.}
\end{abstract}
    


\begin{IEEEkeywords}Virtualization, IaaS, Hy\- pervisors, Performance improvement, Public cloud computing, Ethiopian SMEs, Cloud Computing, Virtualization, Infrastruc\- ture as a Service (IaaS), Performance Optimization, Virtual Machine (VM), Hypervisors, Full Virtualization, Paravirtual\- ization, Hardware\-Assisted Virtualization, Cloud Platforms, Re\- source Utilization, Latency Reduction, Energy Efficiency, High\- Performance Computing (HPC), Open\-Source Solutions, Public Cloud, Ethiopian Small and Medium Enterprises (ESMEs), Amazon EC2, Scalability, Cloud Simulation Tools (CloudSim, GreenCloud, iCanCloud) I~KEYWORDS Cloud computing, Virtualization, IaaS, Hypervisors, Per\- formance improvement, Public cloud computing, Ethiopian SMEs, Infrastructure as a Service (IaaS), Virtual Ma\- chine (VM), Full Virtualization, Paravirtualization, Hardware\- Assisted Virtualization, Cloud Platforms, Resource Utiliza\- tion, Latency Reduction, Energy Efficiency, High\- Performance Computing (HPC), Open\-Source Solutions, Amazon EC2, Scalability, Cloud Simulation Tools (CloudSim, GreenCloud, iCanCloud), Resource Allocation, Latency and Throughput, Open\-Source Platforms, Scalable Cloud Solutions, Cloud In\- frastructure Efficiency, Cost\- Effective Cloud Solutions\end{IEEEkeywords}
%
\IEEEpeerreviewmaketitle

\section{}
\textbf{Index}\textbf{\space }\textbf{Terms}\textbf{{\textemdash}Cloud}\textbf{\space }\textbf{computing,}\textbf{\space }\textbf{Virtualization,}\textbf{\space }\textbf{IaaS,}\textbf{\space }\textbf{Hy-}\textbf{\space }\textbf{pervisors, Performance improvement, Public cloud computing,}\textbf{\space }\textbf{Ethiopian SMEs, Cloud Computing, Virtualization, Infrastruc-}\textbf{\space }\textbf{ture}\textbf{\space }\textbf{as}\textbf{\space }\textbf{a}\textbf{\space }\textbf{Service}\textbf{\space }\textbf{(IaaS),}\textbf{\space }\textbf{Performance}\textbf{\space }\textbf{Optimization,}\textbf{\space }\textbf{Virtual}\textbf{\space }\textbf{Machine}\textbf{\space }\textbf{(VM),}\textbf{\space }\textbf{Hypervisors,}\textbf{\space }\textbf{Full}\textbf{\space }\textbf{Virtualization,}\textbf{\space }\textbf{Paravirtual-}\textbf{\space }\textbf{ization, Hardware-Assisted Virtualization, Cloud Platforms, Re-}\textbf{\space }\textbf{source Utilization, Latency Reduction, Energy Efficiency, High-}\textbf{\space }\textbf{Performance Computing (HPC), Open-Source Solutions, Public}\textbf{\space }\textbf{Cloud,}\textbf{\space }\textbf{Ethiopian}\textbf{\space }\textbf{Small}\textbf{\space }\textbf{and}\textbf{\space }\textbf{Medium}\textbf{\space }\textbf{Enterprises}\textbf{\space }\textbf{(ESMEs),}\textbf{\space }\textbf{Amazon EC2, Scalability, Cloud Simulation Tools (CloudSim,}\textbf{\space }\textbf{GreenCloud,} \textbf{iCanCloud).}
    
\section{KEYWORDS}
Cloud computing, Virtualization, IaaS, Hypervisors, Per- formance improvement, Public cloud computing, Ethiopian SMEs, Infrastructure as a Service (IaaS), Virtual Ma- chine (VM), Full Virtualization, Paravirtualization, Hardware- Assisted Virtualization, Cloud Platforms, Resource Utiliza- tion, Latency Reduction, Energy Efficiency, High-Performance Computing (HPC), Open-Source Solutions, Amazon EC2, Scalability, Cloud Simulation Tools (CloudSim, GreenCloud, iCanCloud), Resource Allocation, Latency and Throughput, Open-Source Platforms, Scalable Cloud Solutions, Cloud In- frastructure Efficiency, Cost-Effective Cloud Solutions.
    
\section{INTRODUCTION}
Cloud computing has emerged as a transformative technol- ogy that offers significant advantages over traditional IT sys- tems. It enables organizations to adopt more flexible IT infras- tructures, reduce operational costs, and enhance accessibility to information. Through cloud computing, organizations can achieve faster data processing and better scalability than with

traditional IT approaches. Additionally, it promotes the growth of an organization's ICT capabilities by allowing applications to be hosted or computing resources to be accessed on remote servers using a pay-as-you-go model \unskip~\cite{2479797:31853180}.

Cloud computing has fundamentally altered the way busi- nesses utilize IT resources, both internally and externally. Despite the potential economic benefits offered by public cloud services, many organizations remain hesitant to migrate their IT infrastructure to external environments. Private clouds, however, are favored in industries with stringent regulatory re- quirements, such as finance, healthcare, and scientific research. Organizations seeking to maintain control over legacy applica- tions also find private clouds appealing. However, with proper security measures in place, the majority of workloads can be securely migrated to public clouds, which offer availability and business continuity comparable to private clouds \unskip~\cite{2479797:31853184}.

Cloud computing can be deployed in various models, but the four primary deployment models include Private Cloud, Public Cloud, Hybrid Cloud, and Community Cloud.
\bgroup
\fixFloatSize{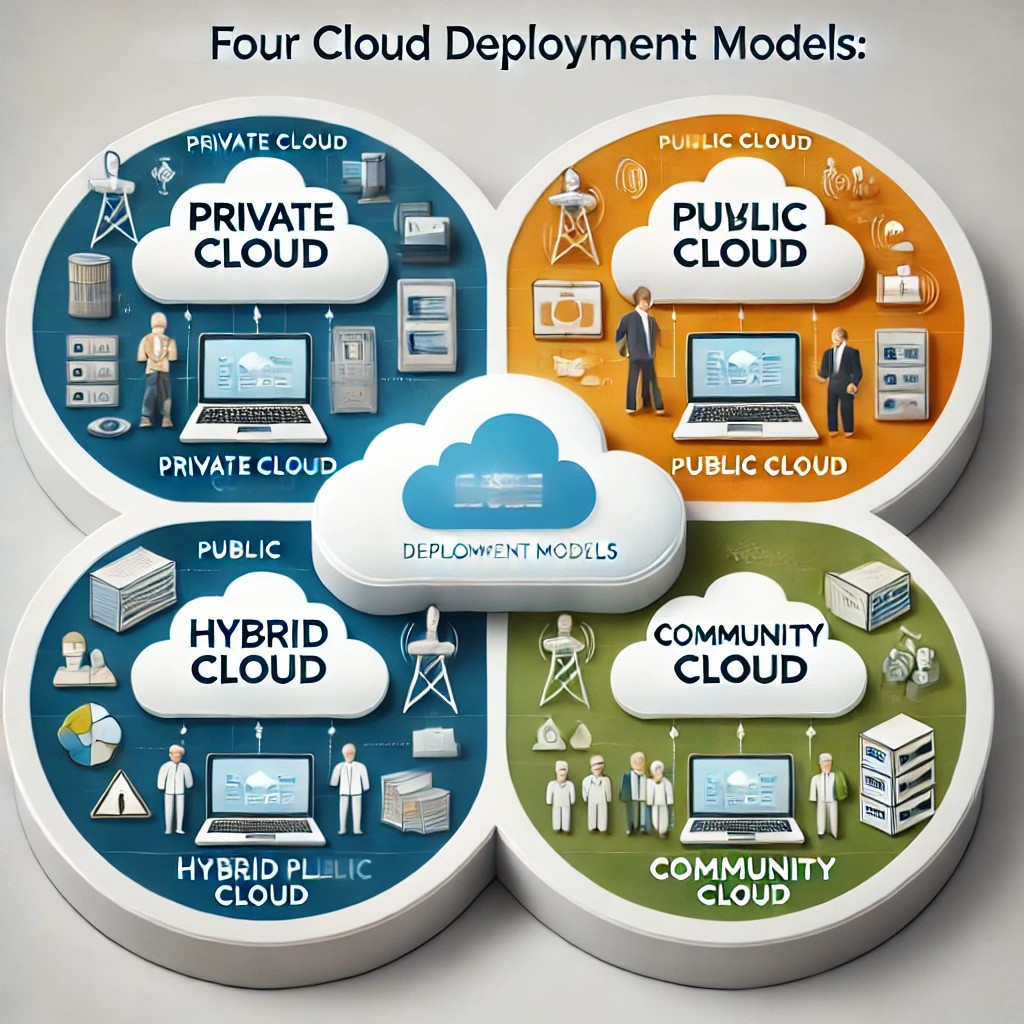}
\begin{figure*}[!htbp]
\centering \makeatletter\IfFileExists{images/c323344b-4174-435b-88b1-0924baaa1916image1.jpeg}{\includegraphics{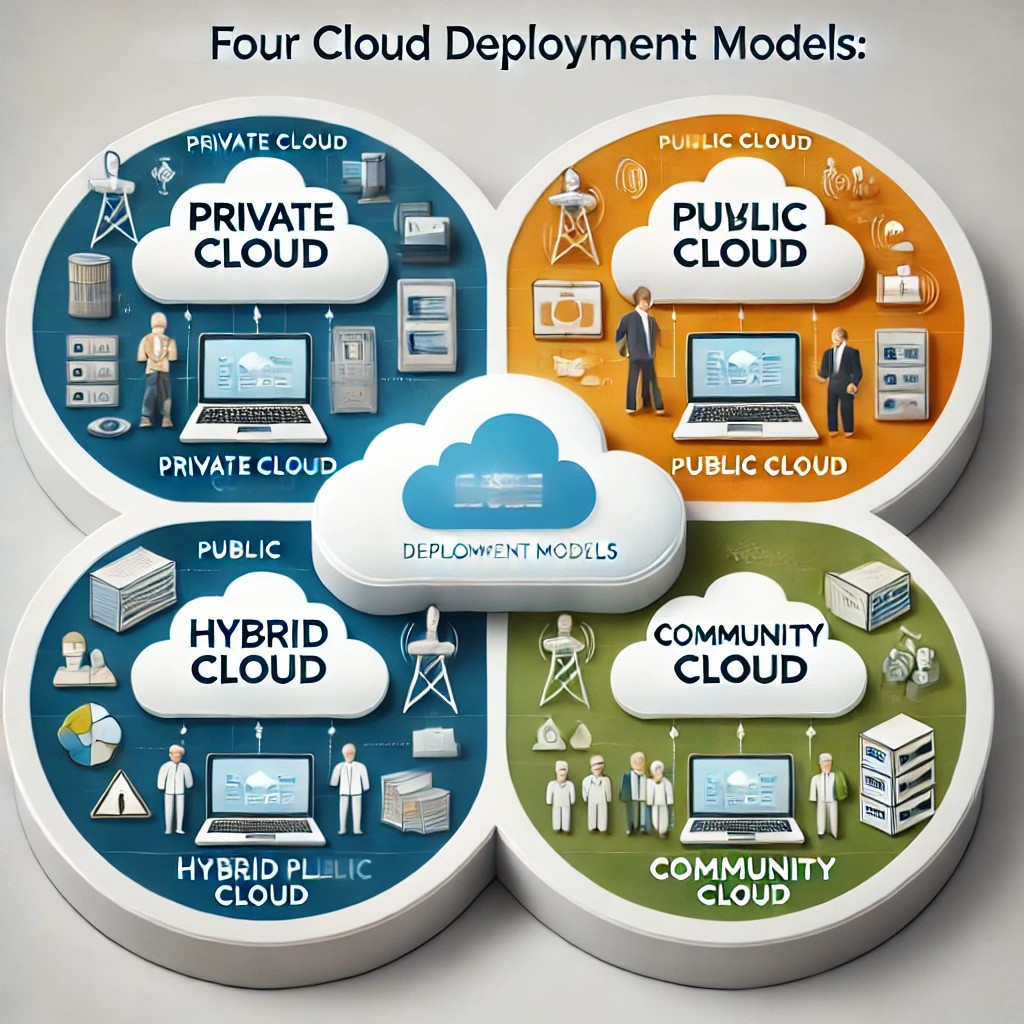}}{\includegraphics{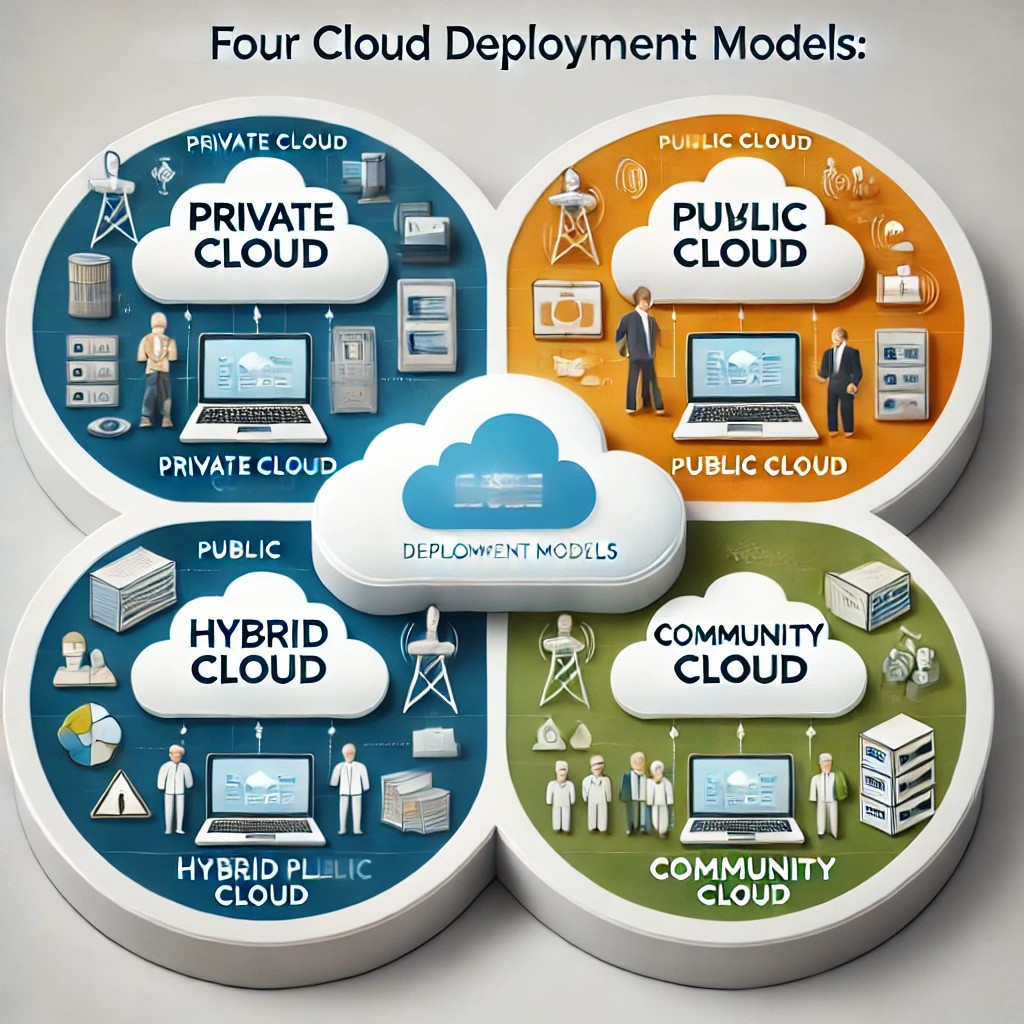}}
\makeatother 
\caption{{1.}}
\label{figure-2b1e0f29c96f41c3b550aab0df4bf9e3}
\end{figure*}
\egroup

\section{Private Cloud: Often referred to as an internal}
or corporate cloud, this model provides resources and data storage to a limited number of hosted services. Private clouds are managed and operated behind an organization's firewall, offering greater control over data \unskip~\cite{2479797:31853182}.

2. Community Cloud: This model is shared among orga- nizations with similar concerns, such as security requirements, policy compliance, or mission alignment. It serves a specific community of users \unskip~\cite{2479797:31853191}.

3. Public Cloud: In this model, cloud services are made available to the general public via the internet. Public clouds can be owned and managed by governments, academic institutions, or commercial entities, providing resources for a wide range of purposes \unskip~\cite{2479797:31853190}.

4. Hybrid Cloud: A hybrid cloud combines multiple cloud deployment models, allowing an organization to manage some resources in-house while utilizing external cloud services for other needs. This model offers the advantages of both public and private clouds.

III. LITERATURE  REVIEW

Efficient Live Wide Area VM Migration with IP Address Change Using Type II Hypervisor

This paper\unskip~\cite{2479797:31853187} presents a solution for virtual machine (VM) migration across a Wide Area Network (WAN). The proposed method allows for VM migration from one host server to another, with the key feature being the ability to change the VM's IP address during the migration process. The solution is divided into three major components: runtime state migration, persistent state migration, and network connection handling. The runtime state migration is executed using VirtualBox's teleportation function, which divides the migration process into stages, each executed at different points. Network connec- tions are maintained through packet redirection and Dynamic DNS. This method addresses common cloud maintenance tasks such as hardware upgrades, scheduled power outages, and server relocation. However, in a cloud environment where VMs are client-facing, maintaining service level agreements (SLA) becomes critical. The inherent downtime required by this method poses potential challenges, especially when sen- sitive processes are running on the VMs. Additionally, this approach does not account for security concerns, such as packet encryption, which is a significant limitation. Although the solution provides a viable method for migration, it results in service disruption and downtime.

Trusted Computing and Secure Virtualization in Cloud Computing

This paper \unskip~\cite{2479797:31853179} explores trusted computing in the context of public cloud environments, using virtualization as a key security enabler. The authors propose a protocol designed to establish trust when launching virtual machines (VMs) in a public cloud setting. The protocol ensures that users can verify the integrity of a VM when it is powered on or launched. The focus is on securing the VM lifecycle, starting from the launch phase, which the authors identify as critical for building trust in cloud environments. A trusted launch protocol is introduced, enabling users to verify the integrity of the VMs

throughout their lifecycle. The paper highlights the impor- tance of building security into the hypervisor layer in cloud computing environments, allowing for the design of secure protocols at the virtualization stage. However, while the paper suggests that the proposed protocol does not negatively impact cloud performance, it lacks practical implementation details and results, leaving some questions unanswered regarding its real-world applicability.

Design and Development of Virtualized Hybrid Architecture for Cloud Computing

Managing data centers and cloud databases, such as Cloud DBMS, can be costly and energy-intensive. This paper \unskip~\cite{2479797:31853181} proposes a hybrid architecture that integrates traditional DBMS tools with virtualized environments, such as MapRe- duce and parallel DBMS. A key challenge identified in cloud computing environments is the heterogeneous nature of clouds, which may not support a wide variety of applications or protocols at the hypervisor level. This paper builds upon the functionality of MapReduce to address these limitations and extends the work by integrating different application environments into a unified architecture.

Dynamic Resource and Energy-Aware Scheduling of MapReduce Jobs and Virtual Machines in Cloud Datacenters MapReduce is a widely adopted programming model for processing large-scale data across distributed environments, particularly in heterogeneous hardware settings. This pa- per \unskip~\cite{2479797:31853183} addresses the performance and energy-efficiency challenges in cloud datacenters by designing a dynamic scheduler for MapReduce jobs and virtual machines (VMs). The proposed scheduler aims to optimize resource alloca- tion by assigning tasks to nodes in a way that minimizes their overall runtime without impacting tasks that are already running. Additionally, the algorithm selects nodes based on compatibility with existing workloads. The scheduler is also energy-aware, dynamically rebalancing VMs based on their resource utilization to minimize the number of active physical machines. This solution contributes to SLA-aware scheduling while simultaneously improving the energy efficiency of cloud

datacenters.
    
\section{PROBLEM STATEMENT}
Cloud computing (CC) has been the focus of numerous research studies over the past decade, and many have identified various \textbf{performance issues }within this innovative computing model. Researchers widely agree that cloud computing holds immense potential for organizations, as it can significantly reduce hardware and software costs, minimize energy con- sumption, and make server usage far more efficient. However, in practice, \textbf{cloud}\textbf{\space }\textbf{computing}\textbf{\space }does not consistently deliver on these promises, leading to discrepancies between theo- retical advantages and real-world performance. For instance, according to data from the \textbf{Dell}\textbf{\space }\textbf{IT}\textbf{\space }\textbf{Group}, a large per- centage of servers worldwide remain \textbf{underutilized}\textbf{\space }despite the increasing demand for cloud services. It is reported that approximately three-quarters of the global working servers have not exceeded \textbf{20\%}\textbf{\space }\textbf{processor} \textbf{utilization}, resulting in

significant performance inefficiencies across various cloud environments, particularly in the \textbf{Infrastructure as a Service}\textbf{\space }\textbf{(IaaS)}\textbf{\space }\textbf{model} \textbf{\space }\unskip~\cite{2479797:31853185}.

\bgroup
\fixFloatSize{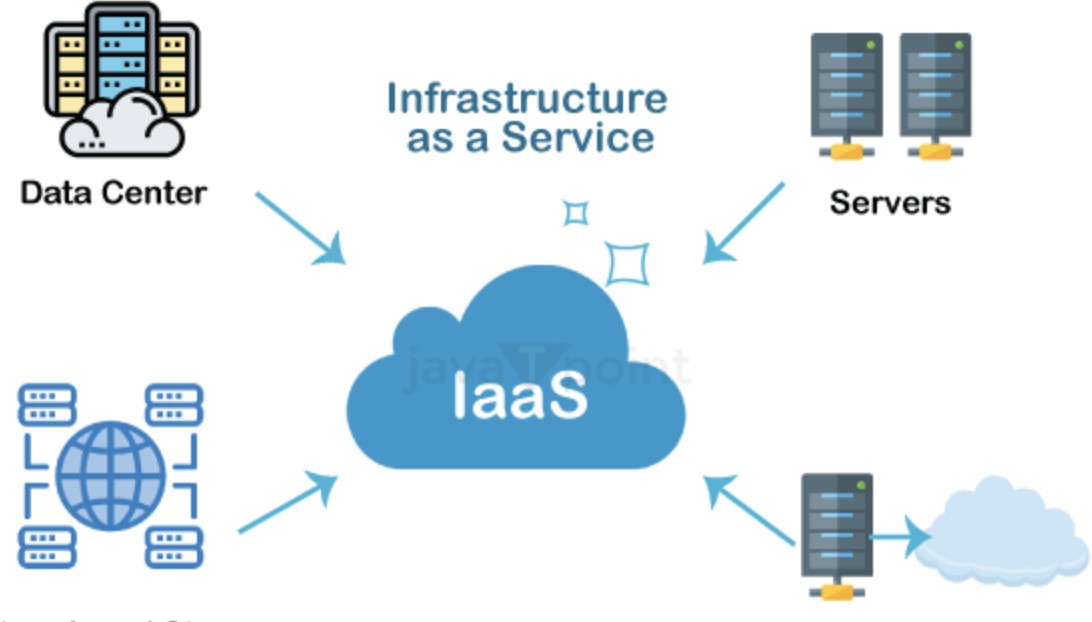}
\begin{figure*}[!htbp]
\centering \makeatletter\IfFileExists{images/c323344b-4174-435b-88b1-0924baaa1916image2.jpeg}{\includegraphics{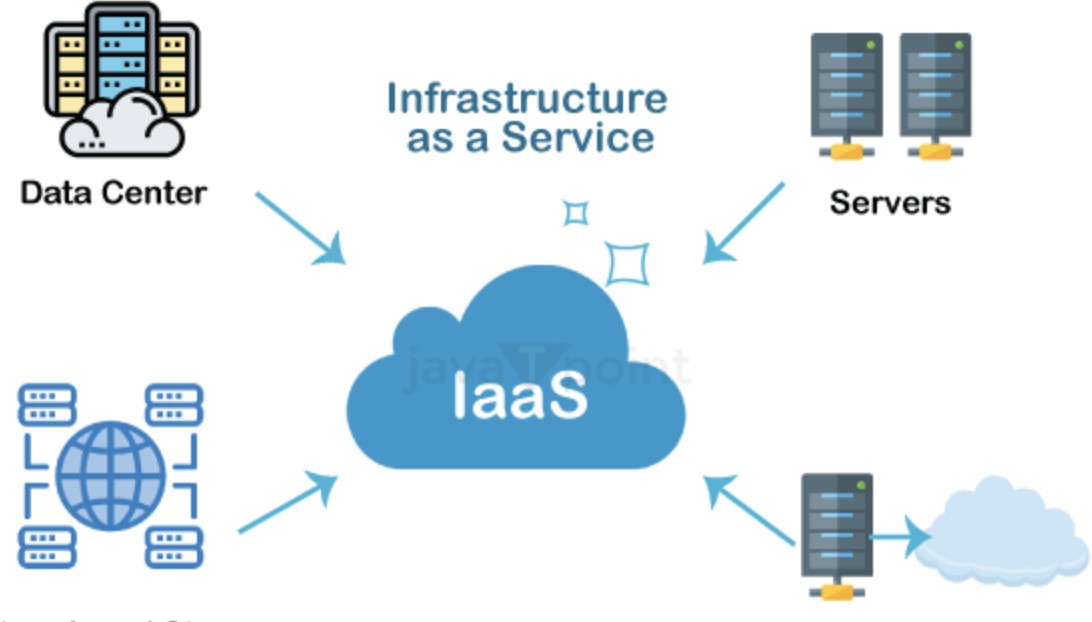}}{\includegraphics{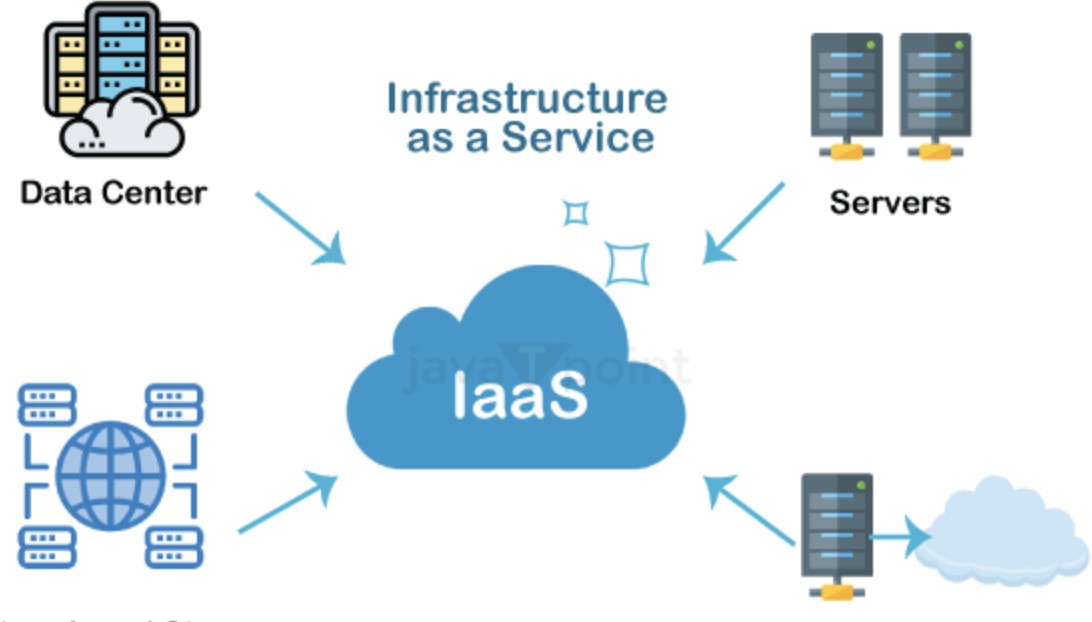}}
\makeatother 
\caption{{2. Illustration of the IaaS model highlighting the infrastructure's relationship with data centers, servers, and cloud services.}}
\label{figure-884e764051de4ebd85b5eff8f7d396f6}
\end{figure*}
\egroup
This underutilization not only highlights the inefficiency in resource allocation but also points to the need for more refined approaches to cloud infrastructure management.

One of the key factors contributing to these \textbf{performance}\textbf{\space }\textbf{inefficiencies}\textbf{\space }is the \textbf{virtualization}\textbf{\space }\textbf{technology}\textbf{\space }employed within cloud computing environments. While \textbf{virtualization}\textbf{\space }is heralded as a core enabling technology for cloud com- puting, playing a pivotal role in resource management and allocation, it also introduces several \textbf{performance trade-offs}\textbf{\space }that can undermine the overall efficiency of cloud systems. Virtualization is regarded as one of the ten key characteristics of cloud computing \unskip~\cite{2479797:31853192}, yet its reliance on sharing physical resources, such as \textbf{CPU cores }and \textbf{network interfaces}, among multiple \textbf{virtual}\textbf{\space }\textbf{machines}\textbf{\space }\textbf{(VMs)}{\textemdash}often belonging to dif- ferent users{\textemdash}leads to notable performance degradation. This \textbf{resource}\textbf{ sharing }frequently results in \textbf{processing}\textbf{\space }\textbf{latency}, \textbf{packet loss}, and a noticeable reduction in \textbf{network through-} \textbf{put}, all of which can significantly hinder the performance of cloud applications and services, especially under high workloads.

Moreover, the reliance on virtualization in cloud computing environments presents additional challenges when striving for \textbf{high-performance computing (HPC) capabilities}. Although \textbf{virtualization }allows for better utilization of existing hard- ware, it struggles to maximize \textbf{throughput }while minimizing the losses associated with \textbf{CPU}\textbf{\space }and \textbf{I/O}\textbf{\space }\textbf{efficiency}. The inherent \textbf{trade-offs }in current virtualization technologies limit cloud environments' ability to fully leverage \textbf{HPC }and other performance-critical applications. This constraint is partic- ularly evident when cloud infrastructure needs to support \textbf{demanding} \textbf{workloads}, where latency and throughput are crucial to the overall performance of the system.

Ultimately, many of the \textbf{performance limitations }observed in cloud computing can be traced back to the fundamental characteristics of the \textbf{virtualization}\textbf{\space }\textbf{technologies} \textbf{\space }and \textbf{hyper-}

\textbf{visor types }that form the backbone of various infrastructure configurations. Each \textbf{virtualization technique}, whether it be \textbf{full virtualization}, \textbf{paravirtualization}, or \textbf{hardware-assisted}\textbf{\space }\textbf{virtualization}, comes with its own set of advantages and disadvantages. However, most of these techniques introduce performance trade-offs that have yet to be fully addressed, leading to \textbf{suboptimal}\textbf{\space }\textbf{cloud}\textbf{\space }\textbf{performance}\textbf{\space }across a wide range of use cases. The challenge of balancing \textbf{resource}\textbf{\space }\textbf{efficiency }with the performance requirements of cloud appli- cations remains one of the most pressing issues in the ongoing development of \textbf{cloud}\textbf{\space }\textbf{computing} \textbf{\space }infrastructure.

\subsection{RESEARCH OBJECTIVES}Cloud computing heavily relies on \textbf{virtualization}\textbf{\space }as an essential enabling technology. Over the years, advancements in processor technology have significantly facilitated the develop- ment and evolution of \textbf{virtualization}, enabling more efficient resource utilization in cloud environments. However, despite these technological advancements, there is still a pressing need to develop more \textbf{effective virtualization techniques }that are specifically tailored for \textbf{IaaS (Infrastructure as a Service)}\textbf{\space }cloud computing environments. The following objectives are categorized into general and specific, outlining the primary goals of this paper in enhancing virtualization for \textbf{IaaS cloud} \textbf{infrastructure}.
    
\section{GENERAL OBJECTIVE:}
The primary objective of this paper is to \textbf{design}\textbf{\space }and \textbf{implement}\textbf{\space }an effective \textbf{virtualization}\textbf{\space }\textbf{technique}\textbf{\space }that will improve the overall performance and efficiency of \textbf{IaaS cloud}\textbf{\space }\textbf{computing infrastructure}. The proposed technique aims to address and overcome the existing limitations, inefficiencies, and \textbf{performance bottlenecks }inherent in current virtualiza- tion implementations. By focusing on these challenges, the resulting solution is expected to either \textbf{enhance }the overall performance of cloud infrastructure or significantly reduce the overhead and resource consumption associated with \textbf{enhance-}\textbf{ ments }made to \textbf{cloud computing technology}. This solution will enable more efficient resource management and better scalability for \textbf{IaaS} \textbf{providers}.
    
\section{SPECIFIC OBJECTIVES:}
In order to develop the most \textbf{appropriate}\textbf{\space }\textbf{and}\textbf{\space }\textbf{high-}\textbf{ performance virtualization technique }for \textbf{IaaS cloud com-}\textbf{\space }\textbf{puting}, the following specific objectives will guide the \textbf{re-}\textbf{\space }\textbf{search}, \textbf{design}, and \textbf{development} \textbf{\space }process:Virtualization and Cloud Computing PerformanceModeling:

The research will involve an in-depth, \textbf{comprehensive un-}\textbf{ derstanding }of the various ways \textbf{virtualization technology}\textbf{\space }creates, manages, and optimizes \textbf{virtual}\textbf{\space }\textbf{machines}\textbf{\space }\textbf{(VMs)} \textbf{\space }within cloud computing environments. A key focus will be placed on identifying and analyzing the techniques that model \textbf{performance improvements }associated with \textbf{virtualization}. This will involve studying the impact of different virtualization

techniques on the overall \textbf{performance}, \textbf{resource allocation}, and \textbf{latency} \textbf{\space }within cloud infrastructure.CarefulSelectionofVirtualizationApproaches:

The paper will meticulously examine and select the \textbf{optimal}\textbf{ approaches }to \textbf{virtualization }based on their effectiveness in improving \textbf{performance}\textbf{\space }and reducing \textbf{resource} \textbf{overhead}. Specifically, careful attention will be given to the following critical aspects of virtualization:approaches will be thor-oughly examined to understand their respective strengths and. The goal is to select theapproachthatmaximizesandwhileminimizingand.

\textbf{virtualization techniques }are applied within real-world cloud environments. Comprehensive documentation will accompany the prototypes, outlining the steps taken during the research, design, and development process. This will include detailed descriptions of the \textbf{algorithms }and \textbf{techniques }used to im- prove \textbf{virtualization} \textbf{efficiency}, as well as the challenges faced and solutions implemented.OptionalObjective:

As an optional component of this research, a \textbf{public cloud}\textbf{\space }\textbf{computing}\textbf{\space }\textbf{design}\textbf{\space }tailored for \textbf{Ethiopian}\textbf{\space }\textbf{Small}\textbf{\space }\textbf{and}\textbf{\space }\textbf{Medium}\textbf{ Enterprises (ESMEs) }may be developed using \textbf{Amazon EC2}\textbf{\space }as the cloud infrastructure. This design will focus specifically on the \textbf{IaaS service model }and will explore how the proposed \textbf{virtualization}\textbf{\space }\textbf{techniques} \textbf{\space }can be applied to provide \textbf{cost-}
    
\section{Pervis\ensuremath{\bullet} ors}
\textbf{Virtual}\textbf{\space }\textbf{Machine}\textbf{\space }\textbf{Managers}\textbf{\space }\textbf{(VMMs)}\textbf{\space }\textbf{or} \textbf{Hy-}
    
\section{Effective, scalable, and high-performance cloud services to}
: The study will evaluate widely used \textbf{VMMs }or \textbf{hypervisors }such as \textbf{Xen}, \textbf{KVM}, and \textbf{VMware}\textbf{ ESXi }\unskip~\cite{2479797:31853178}, comparing their performance in managing \textbf{virtualization re-}\textbf{\space }\textbf{sources}\textbf{\space }and their ability to scale within \textbf{IaaS}\textbf{\space }\textbf{models}. This will involve analyzing the impact of each hypervisor on system- level performance and \textbf{VM}\textbf{\space }\textbf{management} \textbf{\space }efficiency.Evaluation of Cloud Computing Platforms andOpen-SourceSolutions:

The research will assess various \textbf{cloud}\textbf{\space }\textbf{computing}\textbf{\space }\textbf{plat-}\textbf{\space }\textbf{forms}\textbf{\space }and \textbf{open-source}\textbf{\space }\textbf{solutions}, such as \textbf{Eucalyptus}, \textbf{Open-}\textbf{\space }\textbf{Nebula}, \textbf{CloudStack}, and \textbf{OpenStack}. These platforms will be analyzed for their ability to support the proposed \textbf{virtual-}\textbf{ ization techniques }and their impact on the overall \textbf{scalability}, \textbf{resource allocation}, and \textbf{cost-efficiency }of \textbf{cloud infrastruc-} \textbf{ture}.Defining Metrics and Developing PerformanceAnalysisTools:

To ensure accurate evaluation and \textbf{performance}\textbf{\space }\textbf{assess-}\textbf{\space }\textbf{ment}, a set of key \textbf{metrics }will be defined to measure the effectiveness of the proposed virtualization techniques. Metrics such as \textbf{CPU utilization}, \textbf{network throughput}, \textbf{I/O perfor-}\textbf{\space }\textbf{mance}, and \textbf{latency }will be used to assess the improvements made. Additionally, programming code will be developed at the \textbf{hypervisor}\textbf{\space }\textbf{level}\textbf{\space }to evaluate stepwise \textbf{performance} \textbf{ improvements }within cloud infrastructure. These tools will provide a detailed analysis of how virtualization techniques impact system performance.

\bgroup
\fixFloatSize{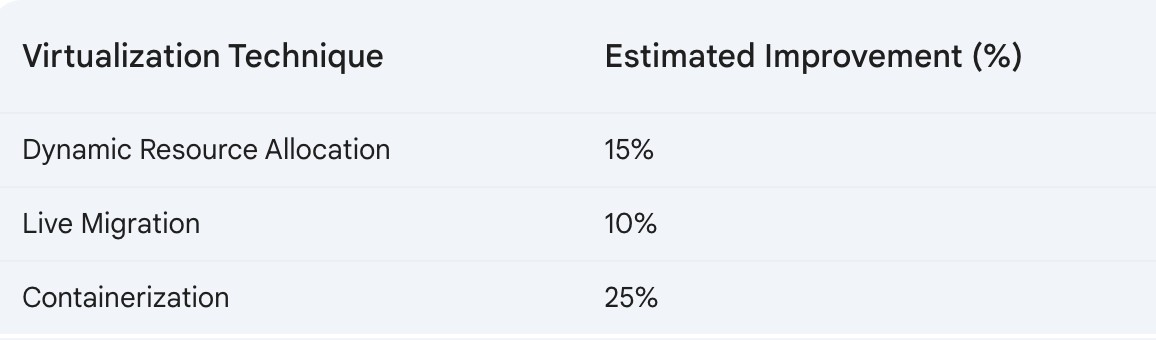}
\begin{figure*}[!htbp]
\centering \makeatletter\IfFileExists{images/c323344b-4174-435b-88b1-0924baaa1916image3.jpeg}{\includegraphics{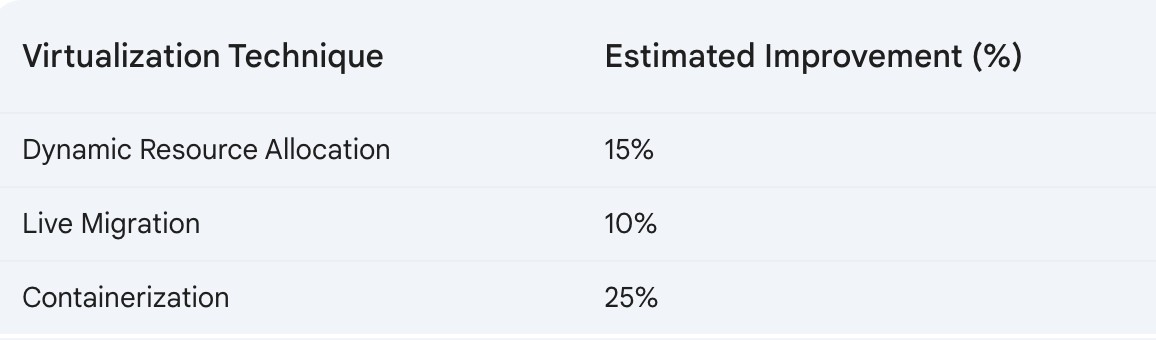}}{\includegraphics{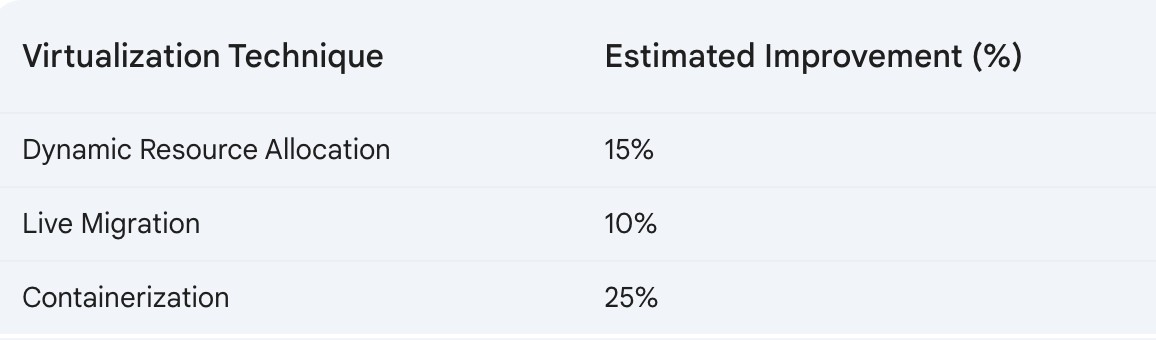}}
\makeatother 
\caption{{3. Estimated Performance Improvements in IaaS Cloud Computing Using Different Virtualization Techniques.PrototypingandDocumentation:}}
\label{figure-0673b3c08b744b66a151bef3056b3f16}
\end{figure*}
\egroup
The study will develop and demonstrate \textbf{prototypes }of the proposed solutions, offering practical examples of how the

small and medium-sized businesses in developing economies.
    
\section{SCOPE}
The scope of this paper is centered on the design, devel- opment, and evaluation of the \textbf{most effective virtualization}\textbf{\space }\textbf{techniques}\textbf{\space }aimed specifically at improving the overall \textbf{perfor-}\textbf{ mance }of \textbf{cloud computing environments}. This research will focus on the \textbf{Infrastructure}\textbf{\space }\textbf{as}\textbf{\space }\textbf{a}\textbf{\space }\textbf{Service}\textbf{\space }\textbf{(IaaS)}\textbf{\space }\textbf{model}, where virtualization plays a critical role in the efficient allocation and management of computing resources. The proposed system will be developed with a focus on \textbf{open-source}\textbf{\space }\textbf{solutions} \textbf{\space }for both \textbf{virtualization }and \textbf{cloud computing }technologies, ensuring that the techniques are accessible, scalable, and adaptable to a wide range of cloud environments.

To fully demonstrate the potential \textbf{performance improve-}\textbf{\space }\textbf{ments}, various \textbf{simulation tools }that are capable of support- ing different cloud computing services may be employed. These simulation tools, such as \textbf{CloudSim}, \textbf{GreenCloud}, or \textbf{iCanCloud}, will be utilized to model the performance of the proposed techniques under varying workload conditions and infrastructure configurations. Through these simulations, this paper will illustrate how the \textbf{virtualization techniques }can enhance \textbf{resource}\textbf{\space }\textbf{utilization}, reduce \textbf{latency}, and improve overall \textbf{system throughput }within IaaS cloud infrastructures. The study will also explore how these techniques can be ap- plied to optimize the performance of existing \textbf{cloud} \textbf{platforms}, such as \textbf{OpenStack}, \textbf{CloudStack}, and \textbf{OpenNebula}.

In addition, while the primary focus is on \textbf{performance}\textbf{\space }\textbf{improvements in cloud computing}, the paper may extend to investigate the broader implications of the proposed solutions, including their impact on \textbf{cost-efficiency}, \textbf{scalability}, and \textbf{resource}\textbf{ management }in various cloud environments. The goal is to create a solution that not only addresses current performance bottlenecks but also provides a framework that can be adapted for future advancements in \textbf{virtualization }and \textbf{cloud} \textbf{technology}.
    
\section{METHODOLOGIES Systematic Literature Review (SLR :}
To gather relevant insights, I~will conduct a systematic literature review (SLR) focusing on the latest findings related

to virtualization technology within cloud computing. This re- view will emphasize how virtualization is evolving beyond its role as merely an enabling technology. Additionally, the study will explore the working principles of different hypervisors (VMMs), including Xen, KVM (Kernel-Based VM Manager), and VMware ESXi. Although these hypervisors have well- defined tasks in managing virtual machines, this review will explore opportunities for performance improvements at the backend level.

Design, Development, and Evaluation Approach:

Based on the findings from the literature review, the next step will involve designing the most effective virtualization performance improvement technique tailored for IaaS cloud computing. This phase will require careful consideration of several key factors in the development of virtualized cloud computing systems, including:VirtualizationApproaches:Fullvirtualization,paravirtualization,andhardware-assistedvirtualization.Virtual Machine Managers (VMMs) or Hypervisors:Xen,KVM,andVMwareESXiCloud Platforms and Open-Source Solutions: Selec-tion of suitable platforms and solutions such as Eucalyptus,OpenNebula,CloudStack,andOpenStack.

The design and development of the proposed virtualization technique may involve variations in performance evaluations due to differences in the internal structures of these open- source solutions. Therefore, I~may use general simulation software, such as CloudSim, Cloud Analyst, Green Cloud, or iCanCloud, which support cloud service providers like Amazon Web Services (AWS), Microsoft, Google, and IBM. Additionally, as an optional task within the scope of this paper, I~will design a Public Cloud Computing solution tailored for Ethiopian Small and Medium Enterprises (ES- MEs), leveraging Amazon EC2 as the infrastructure platform. The deployment type for this design will be Public Cloud

Computing.

\subsection{EXPECTED RESULTS}Based on the proposed virtualization technique and the implementation of performance improvements for \textbf{Infrastruc-}\textbf{\space }\textbf{ture}\textbf{\space }\textbf{as}\textbf{\space }\textbf{a}\textbf{\space }\textbf{Service}\textbf{\space }\textbf{(IaaS)} \textbf{\space }cloud computing environments, several key performance improvements are anticipated:
    
\section{Increased Resource Utilization:}
By optimizing the virtualization techniques (such as \textbf{full}\textbf{\space }\textbf{virtualization}, \textbf{paravirtualization}, and \textbf{hardware-assisted}\textbf{\space }\textbf{virtualization}), the system is expected to demonstrate a sig- nificant increase in resource utilization, particularly regarding \textbf{CPU usage }and \textbf{network throughput}. This will address the current issue of underutilized servers, where a large proportion of global working servers often operate at less than \textbf{20\%} \textbf{capacity}.
    
\section{Reduced Latency:}
The proposed technique aims to reduce the \textbf{latency }associ- ated with virtualization by optimizing \textbf{virtual machine (VM)}\textbf{ management }and resource allocation. This will enable better \textbf{I/O}\textbf{\space }\textbf{performance} \textbf{\space }and improve the response time for cloud

applications hosted in \textbf{IaaS} \textbf{\space }environments, ensuring a smoother user experience for mission-critical applications.
    
\section{Enhanced Scalability:}
The use of \textbf{open-source platforms }such as \textbf{OpenStack }and \textbf{CloudStack }is expected to improve scalability in the cloud infrastructure. The proposed virtualization solution will allow cloud service providers to handle larger workloads without a corresponding increase in \textbf{overhead}\textbf{\space }\textbf{costs}\textbf{\space }or \textbf{resource} \textbf{contention}.
    
\section{Energy Efficiency:}
By dynamically managing resources and optimizing \textbf{vir-}\textbf{\space }\textbf{tual}\textbf{\space }\textbf{machine} \textbf{migrations}, the system is expected to lead to improvements in \textbf{energy efficiency}. This will reduce the total energy consumption of data centers, contributing to more sustainable cloud computing practices.
    
\section{Improved Business Continuity:}
The proposed virtualization techniques will enhance \textbf{system}\textbf{ availability }and \textbf{business continuity}, particularly in \textbf{public}\textbf{\space }\textbf{cloud}\textbf{\space }\textbf{environments}. This is crucial for \textbf{Ethiopian}\textbf{\space }\textbf{Small} \textbf{and Medium Enterprises (ESMEs)}, who will benefit from reliable cloud services tailored to their specific needs.
    
\section{DISCUSSION}
The findings of this research will have profound implica- tions for the design and operation of IaaS cloud infrastructures, particularly in environments where performance and resource management are critical. The proposed virtualization tech- niques, which aim to minimize resource overhead, increase server utilization, and improve overall system performance, are expected to bridge the gap between the current limi- tations of cloud computing and the growing demands for high-performance computing (HPC) applications and other resource-intensive workloads.

One of the primary challenges addressed in this study is the underutilization of server resources that plagues many cloud environments today. Virtualization, while essential to the flexibility of cloud services, often introduces performance trade-offs, including increased latency and reduced throughput. This research focuses on identifying and implementing the most suitable virtualization techniques, such as hardware- assisted virtualization and paravirtualization, to mitigate these issues. By optimizing how virtual machines interact with physical resources, this study provides a clear pathway to improving the efficiency of cloud resource allocation, which is critical for meeting the performance requirements of modern applications.

Additionally, by leveraging open-source platforms in the design and evaluation of the proposed solution, the research ensures that the solution is not only cost-effective but also scalable and widely applicable. This makes the solution acces- sible to cloud providers and businesses of all sizes, including enterprises operating in developing economies. The ability to scale cloud environments efficiently while avoiding the typical trade-offs associated with traditional virtualization techniques has the potential to transform the cloud service industry.

An important focus of this research is its tailored approach to Ethiopian Small and Medium Enterprises (ESMEs). By providing a cloud solution that addresses the specific needs of ESMEs, the research demonstrates the power of virtualization in delivering cost-effective, scalable, and high-performance cloud services. This approach not only highlights the versatil- ity of cloud computing in resource-constrained environments but also offers a blueprint for how other developing markets can benefit from similar cloud infrastructure improvements. As ESMEs increasingly look to cloud technologies to grow their operations, this research positions itself as a valuable contribu- tion to both the cloud computing field and the broader goal of fostering technological development in emerging economies.
    
\section{CONCLUSION}
This paper presents a proposed solution for improving per- formance in \textbf{Infrastructure}\textbf{\space }\textbf{as}\textbf{\space }\textbf{a}\textbf{\space }\textbf{Service}\textbf{\space }\textbf{(IaaS)}\textbf{\space }cloud comput- ing environments through the optimization of \textbf{virtualization}\textbf{\space }\textbf{techniques}. By conducting a comprehensive review of existing literature and employing a structured \textbf{design, development,}\textbf{\space }\textbf{and}\textbf{\space }\textbf{evaluation}\textbf{\space }\textbf{approach}, this research addresses the per- formance limitations that currently hinder cloud computing's full potential. Through the exploration of \textbf{full virtualization}, \textbf{paravirtualization}, and \textbf{hardware-assisted virtualization}, as well as the careful selection of \textbf{hypervisors}\textbf{\space }such as Xen, KVM, and VMware ESXi, this work seeks to minimize \textbf{resource}\textbf{\space }\textbf{underutilization}, \textbf{latency}, and \textbf{network}\textbf{\space }\textbf{throughput}\textbf{\space }\textbf{bottlenecks} \textbf{\space }in IaaS environments.

The proposed design demonstrates that by improving \textbf{hyper-}\textbf{ visor configurations }and refining \textbf{virtualization techniques}, it is possible to significantly enhance cloud infrastructure performance. Furthermore, the application of this virtualiza- tion solution in the context of \textbf{public}\textbf{\space }\textbf{cloud}\textbf{\space }\textbf{computing}\textbf{\space }for \textbf{Ethiopian}\textbf{\space }\textbf{Small}\textbf{\space }\textbf{and}\textbf{\space }\textbf{Medium}\textbf{\space }\textbf{Enterprises}\textbf{\space }\textbf{(ESMEs)} \textbf{\space }showcases its potential for providing scalable and efficient cloud services tailored to the needs of local businesses. Future work may explore further \textbf{performance metrics }and consider the scalability of the solution across broader cloud service environments.

The study has highlighted the \textbf{trade-offs }involved in vari- ous virtualization technologies and demonstrated how careful optimization can yield significant improvements in cloud com- puting environments. By addressing issues related to \textbf{resource}\textbf{\space }\textbf{allocation}, \textbf{hypervisor}\textbf{\space }\textbf{performance}, and \textbf{network}\textbf{\space }\textbf{efficiency}, this work contributes to advancing the capabilities of \textbf{IaaS}\textbf{\space }\textbf{cloud infrastructures}. The adoption of these enhanced vir- tualization techniques can potentially result in \textbf{better}\textbf{\space }\textbf{cost-}\textbf{\space }\textbf{efficiency}, \textbf{higher}\textbf{\space }\textbf{resource}\textbf{\space }\textbf{utilization}, and \textbf{improved}\textbf{\space }\textbf{service} \textbf{quality}, making cloud computing a more viable option for businesses of all sizes.

Moreover, the design and implementation of a \textbf{public cloud}\textbf{ computing }solution for \textbf{Ethiopian Small and Medium En-}\textbf{\space }\textbf{terprises} \textbf{ (ESMEs) }serves as a practical application of the proposed techniques. It demonstrates the flexibility of the solution and its adaptability to specific market needs. The potential to integrate these virtualization improvements into

global cloud platforms such as \textbf{Amazon}\textbf{\space }\textbf{EC2}\textbf{\space }opens further opportunities for \textbf{cloud}\textbf{\space }\textbf{service}\textbf{\space }\textbf{providers}\textbf{\space }to optimize their infrastructure and offer enhanced services to their customers. Future research can build on this work by exploring \textbf{au-}\textbf{\space }\textbf{tomated}\textbf{\space }\textbf{optimization}\textbf{\space }\textbf{tools}\textbf{\space }for virtualization management and expanding the scope of performance evaluation across diverse cloud platforms. Additionally, investigating \textbf{security}\textbf{\space }\textbf{implications}\textbf{\space }and \textbf{scalability} \textbf{\space }in multi-cloud environments will offer deeper insights into how these virtualization techniques

can be universally applied.

\section*{ACKNOWLEDGEMENTS}I would like to express my sincere gratitude to my advisor, Mr. Menore T, and my co-advisor, Dr. Yalemzewd, for their invaluable guidance and support throughout this research. I~am also grateful to Fitsum Assamnew, Computer Engineer, Com- puter Security Expert, and AI Expert, for his crucial advice and program supervision. Their insights and expertise have greatly contributed to the success of this work. Lastly, I~would like to thank the Department of Electrical and Computer Engineering at AKU University for providing the resources and academic environment that enabled this research to flourish.


%

\bibliographystyle{IEEEtran}

\bibliography{\jobname}
\vfill
\end{document}